\documentclass[10pt,journal]{IEEEtran}
\usepackage{amsmath}
\usepackage{amsfonts}
\usepackage{graphicx}
\usepackage{float}
\usepackage{tabularx}
\usepackage{amssymb}
\usepackage{pdflscape}  
\usepackage{setspace} 
\usepackage{float} 
\usepackage{color}      
\usepackage{graphicx}        
\usepackage{graphics} 
\usepackage{subfig}                             
\usepackage{epsfig} 
\usepackage{times} 
\usepackage[hyphens]{url}  
\usepackage{comment}
\usepackage{algorithmicx,algorithm}
\usepackage{bbm}
\usepackage{algorithm} 
\usepackage{algpseudocode} 
\usepackage{algorithmicx}
\usepackage{pifont}
\usepackage{cite}
\usepackage{multirow}

\title{Early Detection and Classification of Hidden Contingencies in Modern Power Systems: A Learning-based Stochastic Hybrid System Approach\thanks{This work is supported in part by National Science Foundation under Grants DMS-2229109.}}
\author{\IEEEauthorblockN{Erfan Mehdipour Abadi, Hamid Varmazyari, Masoud H. Nazari~\IEEEmembership{Senior Member,~IEEE,}}
}
\date{}

\begin{document}
\maketitle
\begin{abstract} 

This paper introduces a novel learning-based Stochastic Hybrid System (LSHS) approach for detecting and classifying various contingencies in modern power systems. Specifically, the proposed method is capable of identifying hidden contingencies that cannot be captured by existing sensing and monitoring systems, such as failures in protection systems or line outages in distribution networks. The LSHS approach detects contingencies by analyzing system outputs and behaviors. It then categorizes them based on their impact on the SHS model into physical, control network, and measurement contingencies. The stochastic hybrid system (SHS) model is further extended into an advanced closed-loop framework incorporating both system dynamics and observer-based state estimation error dynamics. Machine learning methods within the LSHS framework are employed for contingency classification and rapid detection. The practicality and effectiveness of the proposed methodology are validated through simulations on an enhanced IEEE-33 bus system. The results demonstrate that the LSHS framework significantly improves the accuracy and speed of contingency detection compared to state-of-the-art methods, offering a promising solution for enhancing power system contingency detection.

\end{abstract}

\section{Introduction}
\label{Sec1}

Modern power systems (MPS) are complex systems that involve various components and subsystems. For instance, the Midcontinent Independent System Operator (MISO) transmission grid has over 16,000 substations, 25,000 buses, and 38,000 lines \cite{7592475}. 
Power systems are traditionally designed for $N-1$ reliability, meaning they are resilient to the failure of a single component. However, simultaneous occurrences of two or more contingencies can overwhelm the system and lead to widespread blackouts \cite{climategov, HF3}. This underscores the critical importance of detecting contingencies at an early stage.

Certain contingencies, particularly those that cannot be directly measured, remain hidden during normal operations but may become evident when the system is under stress \cite{HF1, HF2, zhai2020identifying}. A notable example is the 2018 Camp Fire in California, triggered by equipment failures from PG\&E. The Camp Fire caused at least 85 civilian fatalities and led to the destruction of 153,336 acres and 18,804 structures, highlighting the devastating consequences of undetected contingencies.

According to \cite{HF1, ZAKARIYA2023108928, HF5}, hidden failures of the protection systems are the key contributors to the wide-area disturbance. The cause of protection failures could be due to errors in their measurements, setting, communication, loading stress, etc. High-order $N-k$ contingencies represent another type of hidden failure that is often underestimated in traditional screening processes \cite{HF3, che2017screening}.
Another source of hidden failures stems from inadequate measurement capabilities, a common issue in most distribution networks \cite{ostrometzky2019physics, sim2024detection}. This deficiency delays the timely detection of contingencies such as distribution line outages. Additionally, the interconnection of physical and communication networks within MPS makes them prone to cyber-attacks and anomalies that may evade detection by monitoring systems \cite{liu2016cyber, liu2016false}.

The growing complexity of MPS, driven by the integration of renewable energy sources, advanced communication networks, and diverse system components, has heightened the need for effective monitoring to prevent wide-area disturbances and cascading failures by timely detection of contingencies. To enhance this capability, modern methods such as statistical modeling \cite{sta1, sta2, sta3, sta4}, optimization \cite{opt1, opt2, opt3}, numerical \cite{num1, num2, num3}, and AI-based techniques \cite{heidari2022accurate, ZAKARIYA2023108928, Guo2023, learn1}, have been proposed.

However, existing methods for detecting contingencies primarily rely on direct measurement data and/or historical contingency signatures, which limits their effectiveness. For example, statistical methods \cite{sta1, sta2, sta3, sta4} often lack adaptability to new or rare contingencies, as they are typically designed based on pre-existing patterns. Optimization-based methods \cite{opt1, opt2, opt3} face similar challenges, struggling to generalize to novel scenarios or rare events.
Numerical methods \cite{num1, num2, num3} depend heavily on high-quality data, which is often scarce or unavailable in real-world applications. Moreover, these methods are computationally intensive, making them less practical for real-time contingency detection.

In practice, not all contingency signatures are known or measurable directly. As a result, existing methods may fail to detect certain classes of contingencies, particularly hidden contingencies. This underscores the urgent need for a generic and comprehensive monitoring framework capable of identifying various types of contingencies across diverse operating conditions.

The primary motivation for this paper stems from the observation that, while existing monitoring systems may not directly detect all contingencies, these events often induce subtle yet identifiable changes in the system's dynamics.

From a mathematical perspective, contingencies can be modeled as discrete-time stochastic events occurring within the continuous operation of MPS. This dual nature of contingencies—discrete events within a continuous system—forms the basis for developing a robust detection and monitoring framework.

In our earlier work \cite{cd1, cd2}, we developed SHS modeling, estimation and detection methods for MPS under contingencies. In \cite{cd2}, the contingency identification framework is conceptualized as a randomly switched linear system (RSLS). Within this framework, each contingency is distinctively treated as a unique switching scenario. This approach models the occurrence of a contingency as a stochastic switching between two distinct operating scenarios. These switching scenarios are a set of finite and predefined contingencies. The goal of the monitoring system is to promptly detect such contingencies by continuously analyzing the system's dynamics \cite{yin2022joint, cd2}. 

When dealing with large systems and/or a vast number of contingency scenarios, the SHS models proposed in \cite{yin2022joint, cd1, cd2} may struggle to identify contingencies in a timely manner. For example, MISO manages over 11,500 major $N-1$ contingency scenarios with a high probability of occurrence, and its estimator performs contingency analysis every 4 minutes.

This paper builds on our earlier work in \cite{cd1, cd2} by developing a learning-based SHS (LSHS) framework designed to enable faster and more accurate detection and classification of contingencies. This enhanced framework addresses the challenges posed by large-scale systems, improving both efficiency and reliability in contingency management.

We first categorize contingencies based on their impact on the SHS model. Three distinct classes are physical, control network, and measurement contingencies. We then extend the SHS model to a closed-loop system incorporating both observer's state estimator and state feedback control. This refined SHS model is designed to detect different categories of contingencies effectively. Finally, we develop the LSHS framework to detect and characterize contingencies in a short time duration. 

The LSHS algorithm is a model-based classification approach trained on the outputs of the closed-loop SHS model. Thus, it does not require exhaustive data for every possible contingency. Instead, representative samples from each contingency category are sufficient for effective training.

This key feature ensures robustness and adaptability, enabling the LSHS method to reliably detect and classify contingencies, even in the presence of previously unknown scenarios that were not predefined within the RSLS framework. This capability significantly enhances the practicality and resilience of the approach in real-world applications.

The main contributions of this paper are summarized as follows:
\begin{itemize}
\item This paper introduces a novel LSHS framework for the early detection of various types of contingencies, including hidden contingencies, using limited sensing and monitoring. The SHS component models contingencies as discrete events, where changes in the power system's transfer function serve as indicators of a contingency.
\item The learning component of the LSHS framework classifies contingencies into three distinct categories based on their impact on system dynamics. This classification reduces the computational burden of contingency identification in large-scale systems by effectively narrowing the search space, thereby improving the accuracy and speed of detection.
\item The existing SHS model is extended into an advanced LSHS framework by integrating closed-loop system dynamics and state estimation error dynamics. This enhancement enables the detection of contingencies across three domains: physical, control network, and sensing/monitoring.

\end{itemize}

The rest of the paper is organized as follows: 
Section \ref{Sec2} provides the foundation of SHS modeling and outlines the contingency identification process using a time-splitting approach.
Section \ref{Sec3} discusses various types of cyber-physical contingencies and their impacts on SHS dynamics.
Section \ref{Sec4} introduces the novel LSHS approach for early detection and classification of contingencies in cyber-physical power systems.
Section \ref{Sec5} evaluates the performance and effectiveness of the proposed LSHS approach through simulations on the IEEE 33-bus system.
Section \ref{Sec6} summarizes the key findings and conclusions of the study.


\section{MPS Dynamics Modeling and Contingency Identification in the SHS Framework}
\label{Sec2}
In this section, we provide the foundation of the SHS modeling approach based on the state space dynamics of MPS. This model will be used for identification of contingencies \cite{cd1, cd2, yin2022joint}. 
The dynamics of MPS, when operating near specific equilibrium points, are typically modeled using small-signal linearization techniques. However, the occurrence of discrete events within the system can cause significant changes in its structure, behavior, and equilibrium points, necessitating a modeling approach to account for these transitions.

To address this, \cite{yin2022joint} introduces the SHS framework that models discrete events as switches between different operational structures as a sequence of RSLS. 
In \cite{cd2}, the contingency scenarios are assumed to be both known and finite; however, the specific timing of their occurrence and which scenario will take place remain uncertain, making it challenging to detect the underlying cause of the contingency. Within this framework, system operation is divided into fixed time intervals ($\tau$),  with the assumption that the operating scenario remains unchanged during each interval, $t \in [k\tau, (k+1)\tau)$ for $k= 0, 1, 2, ... $.

The problem of contingency identification is addressed in \cite{cd2} using the Stochastic Hybrid System (SHS) framework, where contingencies are modeled as unknown switching events among a predefined set of system scenarios. As a result, contingency identification is formulated as the task of detecting the specific active switching scenario within the system.
In general, the RSLS model is represented as:
\begin{align} \label{GeneralRSLSx}
    \dot{x} &= A(\alpha_k) x + B(\alpha_k) u \\ \label{GeneralRSLSy}
    y &= C(\alpha_k) x
\end{align}
where \eqref{GeneralRSLSx} and \eqref{GeneralRSLSy} describe the dynamics and outputs of SHS, respectively.
$\alpha_k$ is the active operational scenarios of the system in the $k$th time interval, selected from the set of all normal and contingency operational scenarios denoted by $\alpha_k \in \mathcal{S}=\{1,2,3,\dots,m\} $. $A(\alpha)$ represents system matrix, $B(\alpha)$ is the control input, and $C(\alpha)$ is the measurement system. Note that within each time interval, $A(\alpha)$, $B(\alpha)$, and $C(\alpha)$ are assumed to be fixed and known matrices. 

For instance, the coupled dynamics of the electromechanical and electromagnetic components of the system can be modeled using internal states for each dynamic node, as demonstrated in \cite{ilic2013engineering, Ilic}, allowing detailed analysis of resource behavior. 
 When the analysis focuses on the integration of power electronic interface resources, such as distributed energy resources (DERs), the state-space representation can be derived from the KVL and KCL equations in the $dq$ reference frame, as shown in \cite{zhao2020photovoltaic, katiraei2007small}. 
In this work, we assume all the generation units are synchronous generators represented by the swing equation
\begin{equation}
   M_i \dot{\omega}_i+ b_i\omega_i = P^{in}_i - P^{out}_i 
\end{equation}
where $M_i$ and $b_i$ are the inertia and damping factors of the resource, respectively. Also, we have $\dot{\delta}_i=\omega_i$. Thus, the state space of the $i$th dynamic nodes could be represented by choosing $x_i =[\delta_i, \omega_i]$ as the states of the system and using power flow equations for deriving the input $P^{in}_i$ and disturbance $P^{out}_i$ values of the system.

Let $x_i$ be the state variables of the $i^{th}$ dynamic nodes. 
Considering an MPS with $n$ nodes, the desired SHS model is derived by concatenating the states of each dynamic node as
\begin{equation}
\label{RSLSmodel}
\begin{array}{l l}
\frac{d}{dt}\begin{bmatrix}x_1 \\\vdots \\x_n\end{bmatrix}&=
\begin{bmatrix}
A_{11}(\alpha_i) & \cdots & A_{1n}(\alpha_i) \\
\vdots & \ddots & \vdots \\
A_{n1}(\alpha_i) & \cdots & A_{nn}(\alpha_i)
\end{bmatrix}\begin{bmatrix}x_1 \\\vdots \\x_n\end{bmatrix} \\   &+
\begin{bmatrix}
B_{11}(\alpha_i) & \cdots & B_{1n}(\alpha_i) \\
\vdots & \ddots & \vdots \\
B_{n1}(\alpha_i) & \cdots & B_{nn}(\alpha_i)
\end{bmatrix}
\begin{bmatrix}u_1 \\\vdots \\u_n\end{bmatrix}
\end{array}
\end{equation}

\begin{equation}
\label{RSLSmodelOutput}
\begin{array}{l l}	\begin{bmatrix}	y_1 \\	\vdots \\y_n\end{bmatrix} 	&=	
\begin{bmatrix}    C_{1}(\alpha_i) & 0 & \dots & 0 \\
0 & C_{2}(\alpha_i) & \dots & 0 \\
\vdots & \dots & \ddots & \vdots \\
0 & 0 & \dots & C_{n}(\alpha_i)\end{bmatrix}
\begin{bmatrix}	x_1 \\	\vdots \\x_n\end{bmatrix}. \\ 
\end{array}
\end{equation}
where diagonal terms, $A_{ii}$, represent the system matrix of the $i^{th}$ dispatchable node and the off-diagonal terms $A_{ij}$ represent the coupling matrix between nodes $i^{th}$ and $j^{th}$. Also, $B_{ii}$ represents each node's control feedback. $B_{ij}$ represents the control information transmitted from node $i$ to node $j$ for dispatching and optimal control purposes. 

Furthermore, \eqref{RSLSmodelOutput} presents the SHS model for the measurement and monitoring system. In MPS, observability is crucial to state estimation, protection, and control. 
The joint observability of continuous and discrete states can be achieved through a combination of measurement devices in different parts of the system and estimation algorithms. 
The design of \eqref{RSLSmodelOutput} aims to maximize system observability while minimizing the number of required measurements, ensuring an efficient and cost-effective monitoring framework.

To have a comprehensive SHS model, we modify the state space to account for all contingency scenarios.
Although the number of contingency scenarios may be extensive, the stability and resiliency evaluation process can be conducted offline. This offline evaluation ensures that the system's behavior under various contingencies can be comprehensively analyzed in advance, facilitating efficient detection and management in real-time. Furthermore, the proposed framework is designed to be expandable, enabling the inclusion of novel contingencies without requiring modifications to the existing detection procedure.

Following the development of the SHS model, the identification method proposed in \cite{cd2} effectively detects specific contingencies as they occur by employing a search mechanism that compares real-time system measurements against a comprehensive set of pre-computed expected outputs, generated offline. Essentially, the system's real-time outputs are matched to these pre-computed scenarios to determine the most likely current state of the system.

For ongoing system monitoring, we utilize a time-splitting strategy as introduced in \cite{yin2022joint}. Each time interval $t \in [k \tau, (k+1)\tau)$ is divided into two segments:
1) Contingency Identification Segment:
The initial segment $t \in [k \tau, k \tau+\tau_0)$, where $\tau_0 \ll \tau$, is dedicated to identifying contingencies. 2) State Estimation and Mitigation Segment:
The remaining portion of the interval focuses on state estimation and optimal control to mitigate the identified contingencies.

\begin{figure}[t]
  \vspace{-0.5em}  \includegraphics[width=1\linewidth, trim={6cm 6.5cm 5.2cm 3cm},clip]{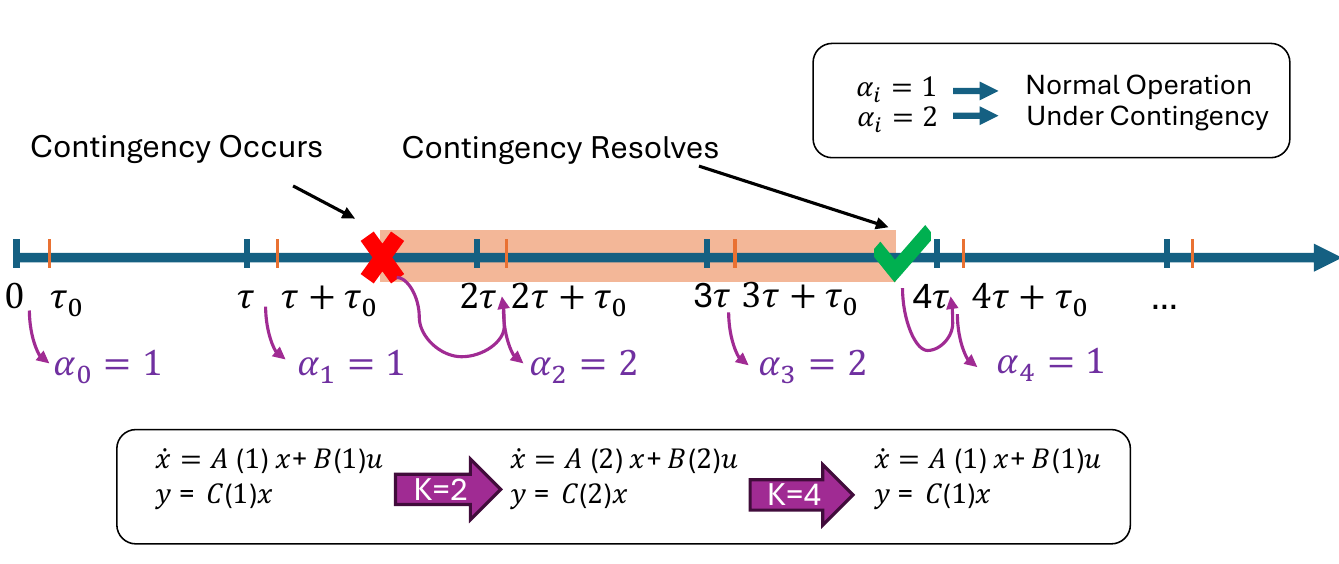}
        \centering
        \captionsetup{width=\linewidth}
        \caption{The time-splitting framework of the RSLS model. }
        \label{Fig_RSLS}
        \vspace{-1.5em}
\end{figure}

This framework, illustrated in Fig. \ref{Fig_RSLS}, ensures timely detection and mitigation of contingencies. For instance:
A contingency occurring at an unknown time within the interval $[\tau, 2\tau)$ is detected during the contingency identification segment $[2\tau, 2\tau+\tau_0)$.
If the contingency persists into subsequent intervals, the detection segment of the fourth interval, 
($[4\tau, 4\tau+\tau_0)$), confirms whether the system has returned to normal operational conditions.

 When a contingency occurs, it typically affects only a specific part of the system, leaving the rest unchanged. As a result, the various switching scenarios of the RSLS that represent these contingencies share common eigenvalues in the matrix 
$A(\alpha)$. This shared eigenvalue structure makes the use of probing input signals essential for effective contingency identification \cite{yin2022joint}.

Since this contingency identification approach examines system behavior in an open-loop format, it is well-suited for detecting contingencies that alter the dynamics captured by 
$A(\alpha)$. However, it has limitations in identifying contingencies that impact the control network (matrix $B(\alpha)$) and/or measurement components (matrix $C(\alpha)$). To address this shortcoming, we propose a closed-loop SHS model in Section \ref{Sec3}, designed to capture a broader range of contingencies, including those affecting control and measurement systems.

Another significant challenge with existing SHS models is their inefficiency in searching large sets of contingency scenarios. In Section \ref{Sec4}, we tackle this problem by introducing the LSHS method. This approach leverages machine learning (ML) techniques to classify contingencies, effectively reducing the search space and improving computational efficiency.

\section{Hidden Contingency Modeling and Classification}
\label{Sec3}

In the following, we introduce a novel categorization of hidden contingencies based on their impacts on the SHS model. We categorize contingencies into three primary groups: physical contingencies ($\mathcal{S}{p}\subset\mathcal{S}$), control network contingencies ($\mathcal{S}{c}\subset\mathcal{S}$), and measurement failures ($\mathcal{S}_{m}\subset\mathcal{S}$). Next, we will detail the mathematical foundations for each category and analyze their potential impacts on MPS stability and functionality.

\subsection{Physical Contingencies}
Physical contingencies are events that directly impacts the physical grid, such as line outages. They primarily  influence the system matrix $A(\alpha)$, as denoted by 
\begin{equation} {A}(\alpha_k)=\sum_{i=1}^m {A}(i) \mathbbm{1}_{{\alpha_k=i}}, \end{equation}
where $\mathbbm{1}_{\alpha}$ is the indicator function of the contingency. In other words, $\mathbbm{1}_{\alpha_k} =1$ in the $k$th time interval if $\alpha_k$ contingency has occurred; and $\mathbbm{1}_{\alpha_k} =0$, otherwise.

For instance, in distribution networks with limited measurement infrastructure, detecting failures such as line outages poses significant challenges \cite{bento2020method}. These contingencies cause changes in the power flow of MPS, leading to changes in the $A(\alpha)$ matrix.
Failures in protection system, such as unexpected changes or malfunctions of switches, often fall under this category \cite{zhao2019review, sodin2023precise}.

Fast detection of physical contingencies is critical for maintaining power balance, reliability, and cost-effectiveness in power systems.
Stability margins and sensitivity analyses can be performed offline for each $A(\alpha)$ before their actual occurrence within the SHS framework, allowing well-prepared mitigation and enhancement strategies. For example, if during a specific contingency, the system matrix $A(\alpha)$ is not \emph{full-rank}, then equation $A(\alpha) x + B(\alpha) u = 0$ lacks a unique solution, indicating the absence of a feasible power flow solution. This scenario necessitates immediate control actions, such as integrating new energy sources to prevent further deficiencies.

\subsection{Control Network Contingencies}
Control network contingencies encompass unforeseen disruptions affecting the functionality of control systems.
These disruptions stem from communication failures, cyber attacks, human errors, or inaccuracies in signal transmission. The impact of control network contingencies on power systems can be significant and can cause major disruptions. Therefore, rapid and accurate detection and identification of these contingencies is crucial for system stability and resilience. 

This type of contingency can appear in various forms such as malfunction, denial of service \cite{liu2013denial}, random operation \cite{en14071989}, delay attacks \cite{rouhani2023adaptive}, false data injection \cite{10155170}, packet loss \cite{cetinkaya2016networked}, or oscillatory behavior \cite{FreqOsc}. They interfere with the transmission of accurate information and control signals across the network, leading to erroneous or absent control commands that compromise system operations.

The impact of control network contingencies is modeled in the SHS framework by modifications to the control matrix $B(\alpha_k)$, expressed as:
\begin{equation} B(\alpha_k)=\sum_{i=1}^m B(i) \mathbbm{1}_{{\alpha_k=i}}. 
\end{equation}

This representation allows for localized failures at node \(i\), impacting \(B_{ii}(\alpha)\) and affecting localized control responses. Similarly, disruptions in communication between nodes \(i\) and \(j\) alter \(B_{ij}(\alpha)\), affecting inter-node control dynamics.

One of the primary challenges is the potential loss of controllability. This risk entails the system’s reduced ability to stabilize or optimize operations during instabilities or other critical situations. To address this, analyzing the controllability matrix is essential:
\begin{equation} \mathcal{C}(\alpha) = \left[B(\alpha) \ \ A(\alpha) B(\alpha) \ \dots \ A(\alpha)^{n-1} B(\alpha)\right]. \end{equation} 
For instance, packet loss in a networked control system disrupts the intended control actions by setting the control input to zero, effectively nullifying the control influence associated with certain system states. We can mathematically represent this condition by setting the corresponding columns in $B((\alpha))$ to zero, directly illustrating that packet loss affects the controllability of the system.

\subsection{Sensing and Monitoring Contingencies}
Contingencies in the sensing and measurement network represent a critical vulnerability within MPS, where inaccurate data fed into the system's state estimation. These failures can arise from malfunctions or failures of sensors and measurement devices, or from intrusions that corrupt the measurement signals used by the state estimation system. Such discrepancies can severely disrupt operational integrity and lead to catastrophic outcomes. Therefore, the effective detection and identification of measurement contingencies are essential to minimizing their potential impact.

Sensor/monitoring contingencies are modeled in the SHS framework by changes in the matrix $C(\alpha_k)$ as follows:
\begin{equation} C(\alpha_k) = \sum_{i=1}^m C(i) \mathbbm{1}_{{\alpha_k=i}}; 
\end{equation}

Observability is a critical aspect of system design that ensures all system states are accurately estimated. The observability matrix is defined as: \begin{equation} \mathcal{O}(\alpha) = \begin{bmatrix} C(\alpha)^T & (C(\alpha)A(\alpha))^T & \dots & (C(\alpha)A(\alpha)^{n-1})^T \end{bmatrix}^T, \end{equation} which serves as a key criterion for the design of the measurement system and the implementation of backup sensors. The system is typically designed to be observable so that an observer can estimate all system states. However, if a sensor fails, its impact must be reassessed by removing the affected sensor's data from the observability analysis. A single sensor failure can render multiple states of the system unobservable, significantly compromising system monitoring capabilities.
Implementing redundancy measures such as backup sensors, data validation algorithms, fault identification and isolation techniques, and Machine learning algorithms are some of the ways to deal with this issue.
Identifying which portion of the system remains observable under each sensor failure scenario is developed by \cite{cd1}. This understanding shapes the detection algorithm's approach, focusing on the subset of the system that remains observable.

\subsection{Closed-Loop SHS model with Observer Error Dynamics}
\label{Sec4}
\begin{figure}[t]
    \includegraphics[width=1\linewidth, trim={1.5cm 2.6cm 3.2cm 2.8cm},clip]{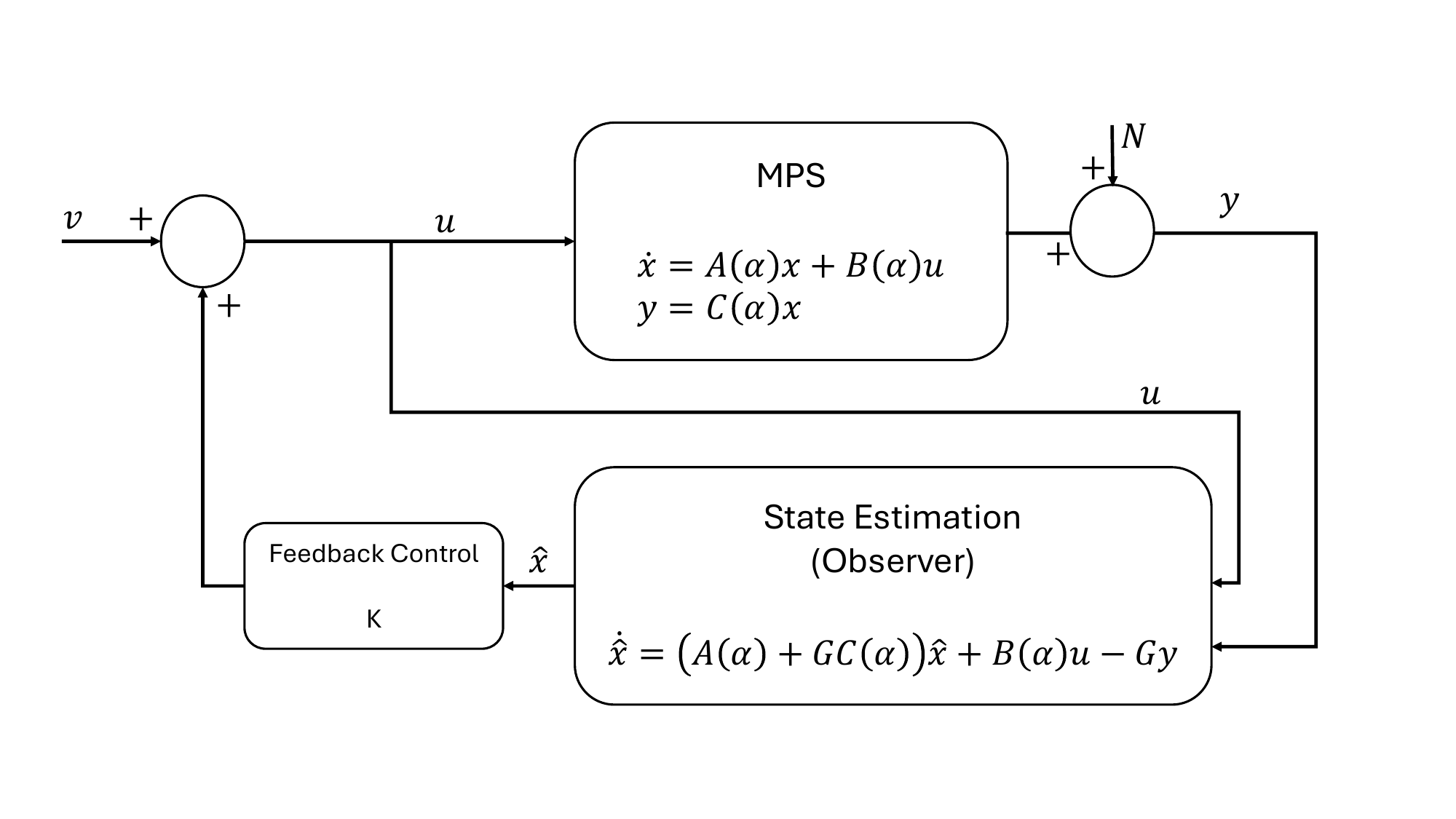}
        \centering
        \captionsetup{width=\linewidth}
        \caption{Feedback control using observer's state estimation \cite{lin2007robust}.}
        \label{Fig_ObserverSchematic}
        \vspace{-1.5em}
\end{figure}

As discussed earlier, the contingency identification in \cite{cd2} performs an open-loop framework where the probe input is applied to the system and switching scenarios are identified based on the system response. We propose incorporating the observer dynamics and feedback control signals alongside system outputs within the SHS framework, based on the system described in \cite{lin2007robust}. This framework is illustrated in Fig. \ref{Fig_ObserverSchematic}.
For each contingency scenario, this system could be modeled by 
\begin{subequations} \label{eq_ObserverStateSpace}
    \begin{align}
        \label{eq_ObserverStateSpaceA} \dot{x} &= A(\alpha)x + B(\alpha) u \\
        \label{eq_ObserverStateSpaceB} y &= C(\alpha) x + N \\
        \label{eq_ObserverStateSpaceC} \dot{\hat{x}} &= (A(\alpha)+GC(\alpha))\hat{x}+B(\alpha)u-Gy \\
        \label{eq_ObserverStateSpaceD} u &= K \hat{x} + v
    \end{align}
\end{subequations}
where $N$ is the independent and zero mean Gaussian measurement noise with variance of $\sigma$. Also, $G$ and $K$ are the system state estimation and feedback controller gains. We assume the observer is designed based on a pole-placement approach similar to \cite{lin2007robust}. Accordingly, we can rewrite \eqref{eq_ObserverStateSpaceA} as
\begin{align}
   \notag \dot{x} &= A(\alpha)x + B(\alpha)(K \hat{x} + v) \\
   \notag \quad   &= A(\alpha)x + B(\alpha) K(x -\tilde{x}) + B(\alpha) v \\
   \quad   &= (A(\alpha)+ B(\alpha) K)x - B(\alpha) K\tilde{x} + B(\alpha) v.
\end{align}
Let $\tilde{x}:= x-\hat{x}$ be the estimation error of the observer. Hence, the estimation error dynamics are 
\begin{align}
    \notag \dot{\tilde{x}} &= \dot{x} - \dot{\hat{x}} \\
    \notag \quad   &= A(\alpha)x + B(\alpha) - \left( (A(\alpha)+GC(\alpha))\hat{x}+B(\alpha)u-Gy \right) \\
    \notag \quad   &= \left( A(\alpha)+GC(\alpha) \right) (x - \hat{x}) + GN\\ 
    \quad  &=\left( A(\alpha)+GC(\alpha) \right) \tilde{x} + GN
\end{align}
Thus, we define the closed-loop SHS model as
\begin{align} \label{eq_AdvanceRSLS}
    \begin{bmatrix}
        \dot{x} \\
        \dot{\tilde{x}}
    \end{bmatrix} 
    &=
    \begin{bmatrix}
        A(\alpha) + B(\alpha)K & -B(\alpha)K \\
        \textbf{0} & A(\alpha)+GC(\alpha)
    \end{bmatrix}
    \begin{bmatrix}
        x \\
        \tilde{x}
    \end{bmatrix} \\ \notag
    &+
    \begin{bmatrix}
        B(\alpha) &  \textbf{0} \\ \textbf{0} & G
    \end{bmatrix}
     \begin{bmatrix} v \\ N
    \end{bmatrix},
\end{align}
where switching of MPS under any category of contingencies are distinguishable by implementing a proper probing input $v$. 

Since we have access to the state estimation information, we define all $\hat{x}$ values in the closed-loop SHS outputs. Therefore the output of the system is defined as
\begin{equation} \label{eq_AdvancedSHSoutput}
    y_c = \begin{bmatrix}
        C(\alpha) & \textbf{0} \\ I & -I
    \end{bmatrix} \begin{bmatrix}
        x \\ \tilde{x}
    \end{bmatrix} + \begin{bmatrix}
        N \\ \textbf{0}
    \end{bmatrix}
\end{equation}
where $y_c =[y, \hat{x}]^T$.

\subsection{Modeling Contingencies on Input and Output Signals}
In this section, we examine contingencies that alter the control input (\( u \)) or measurement output (\( y \)) signals of the system. While some contingencies do not directly modify the state-space matrices (\( A \), \( B \), and \( C \)) in the SHS model, they influence the input and output signals. We demonstrate that such contingencies can still be incorporated into the SHS framework and fall into the classes introduced above by following the procedure outlined in this section.

This type of contingencies that usually stem from cyber-attacks, operational errors, or data handling and processing faults affect the system's operational dynamics by influencing how parameters are interpreted or utilized. For this type of hidden contingencies, we propose considering two equivalent systems based on \eqref{eq_ObserverStateSpace}: the actual system, representing the system under a hidden failure, and the SHS system, which simulates its equivalent behavior as a switching scenario in the SHS model. These systems are defined as follows:
\begin{subequations}\label{H_actual}
\begin{align}
H_{\text{actual}}: & \quad \dot{x} = A_{1}x + B_{1}u, \label{17a}\\
& \quad y = C_{1}x, \label{17b}\\
& \quad \hat{\dot{x}} = (A_{1} + GC_{1})\hat{x} + B_{1}u - Gy.\label{17c},
\end{align}
\end{subequations}

\begin{subequations}\label{H_SHS}
\begin{align}
H_{\text{SHS}}: & \quad \dot{x}' = A_{2}x' + B_{2}u' \\ & \quad y' = C_{2}x' \\ & \quad \hat{\dot{x}}' = (A_{2} + GC_{2})\hat{x}' + B_{2}u' - Gy' ,
\end{align}
\end{subequations}
where the goal is to define the $H_{SHS}$ so that it behaves similar to $H_{actual}$ such that
we assume the state space matrices in $H_{\text{actual}}$ (\(A_1, B_1, C_1\)) remain unchanged, reflecting the normal operation of the system. However, parameters such as \(u, y\) change depending on the contingency. The goal for the \(H_{\text{SHS}}\) is to mirror this behavior without altering \(u', y'\) parameters. Instead, we adjust \(B_2, C_2\) accordingly, ensuring that the dynamics represented by all three equations in \eqref{H_actual} and \eqref{H_SHS} remain consistent across both systems. 

In case of contingencies affecting a control input \(u_i\), the following steps must be carried out:
\begin{enumerate}
    \item Define parameter \(u_i\) under contingency.
    \item Calculate the effect of contingency on \(\dot{x}\), and \(\dot{\hat{x}}\).
    \item Due to the conditions of \(\dot{x}\) = \(\dot{x'}\)  and \(\dot{\hat{x}}\) = \(\dot{\hat{x}}'\), the equality $[B_1]_i u_i = [B_2]_i u'_i$ must hold; where $[B]_i$ represent the $i$th column of $B$.
    \item By defining $[B_2]_i$ = $\frac{[B_1]_i u_i}{u'_i}$, the appropriate $H_{SHS}$ is derived.
 \end{enumerate}
We observe that these hidden contingencies can be incorporated into the SHS framework as part of the control network contingency class by appropriately modifying the matrix $B$.
For example, consider the case of packet loss for one of the control inputs of the system, where \(u_i\) = 0. Thus, \eqref{17a} can be written as $\dot{x} = A_{1}x$. Due to the condition of step 3, we have
\begin{equation}
    A_{1}x = A_{2}x' + B_{2}u'.
\end{equation}
Since this accounts for a control input contingency, we set \(A_{1} = A_{2}\) which leads to \(B_{2} = 0\).
Other types of contingencies may not be as trivial as the packet loss scenario. However, in a normal scenario, we can calculate the effect of the contingencies as a function of the system. 

Similarly, for a contingency that affects the measurement output \(y_i\), similar steps should be followed as in the control input contingency category, with the following modification. Instead of Step 3 in the control input contingency, \(y'_i = y_i\) must be satisfied, resulting in  
\begin{equation}  
    y_i = [C_{2}]^i x',  
\end{equation}  
where \([C_{2}]^i\) represents the \(i\)th row of the matrix \(C_{2}\). Thus, this contingency falls into the sensing and monitoring class of contingencies.

One of the significant challenges in this framework arises from the nature of hidden contingency parameters, which can vary continuously within a specific range, complicating their representation as discrete switching scenarios. To address this issue, we propose a strategy for quantizing the hidden failure parameters. By dividing the continuous range into discrete segments, we define a distinct switching scenario for each segment. This quantization allows for a more structured and manageable modeling approach within the SHS framework, enabling more accurate simulations and analyses of the impacts of these contingencies on the system dynamics. 


\section{LSHS Approach for Contingency Detection and Classification}
In this section, we address the limitations of SHS models as the number of contingency scenarios increases within the system. The contingency identification approach proposed in \cite{cd2} relies on analyzing changes in the system's outputs in response to probing inputs and comparing them with real-time measurements within the time interval $t\in [k\tau,k\tau+\tau_0)$. However, as the number of contingencies in the SHS model grows, two key challenges arise:
1) Increased Computational Burden:
The process of estimating the expected system response across all possible scenarios becomes computationally intensive as the search space expands.
2) Reduced Identification Accuracy:
The accuracy of the identification algorithm diminishes with the growth of the search space, making it more challenging to reliably detect specific contingencies.

\begin{figure}[t]
    \includegraphics[width=1\linewidth, trim={0cm 0cm 0cm 0cm},clip]{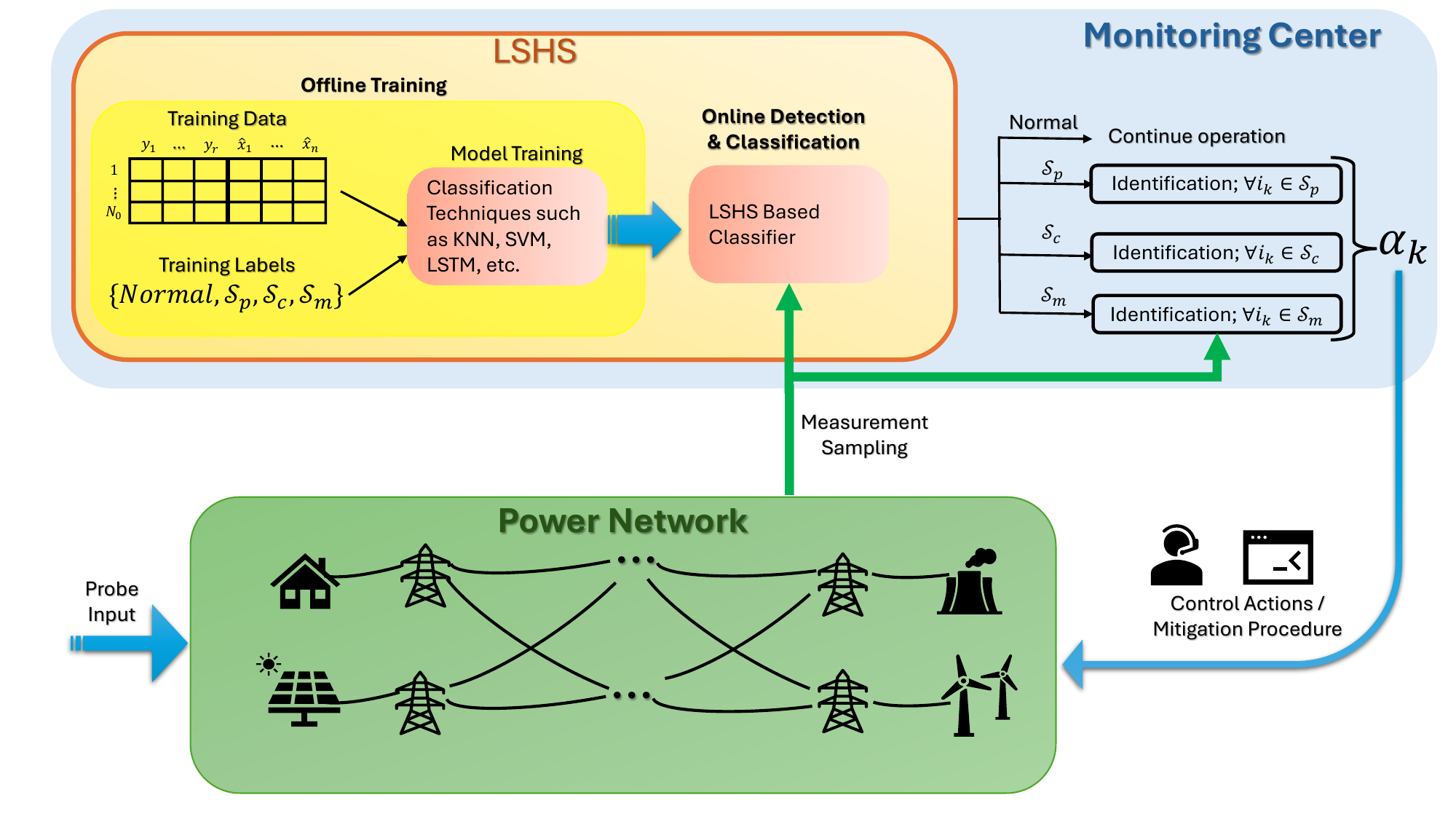}
        \centering
        \captionsetup{width=\linewidth}
        \caption{LSHS-based contingency detection and classification framework. }
        \label{Fig_Classification}
        \vspace{-1em}
\end{figure}

To overcome these challenges, we propose the application of a ML method for contingency classification, as illustrated in Fig. \ref{Fig_Classification}. This approach leverages the capabilities of ML to effectively reduce the search space, enhance computational efficiency, and improve the accuracy of contingency identification in large-scale systems.

The LSHS method is developed based on the closed-loop SHS model described in \eqref{eq_AdvanceRSLS}. The structure of the closed-loop system matrix provides critical insights into the class of contingency. 
Let the eigenvalue sets of $(A(i)+B(i)K)$ and $(A(i)+GC(i))$ be $\Lambda_{1,i}$ and $\Lambda_{2,i}$, respectively.
Control contingencies cause variations only in $\Lambda_{1, i}$. 
Sensor/monitoring contingencies cause variations only in $\Lambda_{2, i}$. But,  Physical contingencies result in changes to both eigenvalue sets.
These physical characteristics are leveraged to label and classify contingency datasets based on their impact on the SHS model. 

The classification process involves evaluating the error values between \( y_c \) obtained from \eqref{eq_AdvancedSHSoutput} under a contingency, and the nominal system output without any measurement noise effect, denoted as \( y_{\text{nom}} \) which is considered as out reference point for system operation. This error is computed as the element-wise difference between the corresponding outputs:

\begin{equation}
e_i(l,k) = \mid[y_c(\tau k +lt_s)]_i - [y_{\text{nom}}(\tau k +lt_s)]_i\mid,
\end{equation}
where \( e_i(k)=[e(1,k), e(2,k), \dots, e(N_0,k)]^T \) is a vector representing the error values for the $i$th output of the system for $i=1,2,\dots, r+n$ assuming 
measurement system $y \in \mathbb{R}^r$ and state estimation $\hat{x}  \in \mathbb{R}^n$ within the time interval \( t \in [k\tau, k\tau + \tau_1) \) where $\tau_1 \leq \tau_0$; and $N_0$ being the number of samples. The concatenation of error values for each output as $e(k) = [e_1(k), e_2(k), \dots, e_{r+n}(k)]$ could be used as the inputs of the time-series-based classification methods such as long short-term memory (LSTM), where each output is representing one of the features of the classification system labeled by the class of contingency under study. However, for conventional classification methods such as K-nearest neighbors (KNN), the sum of error values for each output could be used as the system's features. Since the error values could be small for calculation purposes, we propose using the logarithm of error values instead which would improve the performance of the classification method for this problem. Additionally, a small value of $\epsilon$ is added to the sum of errors to prevent having zero inputs in the logarithm operation. Thus, we have
\begin{equation}
    E_i(k)=log(\sum_{l=1}^{N_0} e_i(l,k) + \epsilon)
\end{equation}
as the aggregation of the error over the $k$th interval.
For classification purposes, the aggregated error \( E_i(k) \) serves as a critical feature to distinguish between different scenarios and  $E(k) = [E_1(k), E_2(k), \dots, E_{r+n}(k)]$ is used as the inputs of the classification procedure. 

The classification process is divided into two stages as demonstrated in Fig \ref{Fig_Classification}: 1) Offline Training: During this phase, a comprehensive dataset is generated by simulating various contingency scenarios. The aggregated error values, computed over the specified intervals, are used to train the ML model. This offline stage leverages the labeled dataset to learn the patterns and relationships between the contingencies and their corresponding error characteristics, ensuring the model achieves high accuracy in identifying and classifying different scenarios. 2) Online Classification: Once the model is trained, it is deployed for real-time contingency identification. In this phase, inputs are processed sequentially. For each new input, the system evaluates the error \( e(k) \), computes the aggregated error \( E(k) \), and utilizes the trained model to classify the contingency. This online approach ensures efficient and accurate real-time decision-making, enabling prompt identification of contingencies as they occur.

Various ML algorithms, including LSTM networks, KNNs, and support vector machines (SVM), would be evaluated for classification.
The LSHS method demonstrates the capability to detect hidden contingencies by utilizing the SHS model outputs, as described in \eqref{eq_AdvancedSHSoutput}. This approach effectively enhances contingency detection and classification, particularly in complex system scenarios.

\section{Simulation}
\label{Sec5}
\begin{figure}[t]
    \includegraphics[width=1\linewidth, trim={3.5cm 1cm 1cm 0cm},clip]{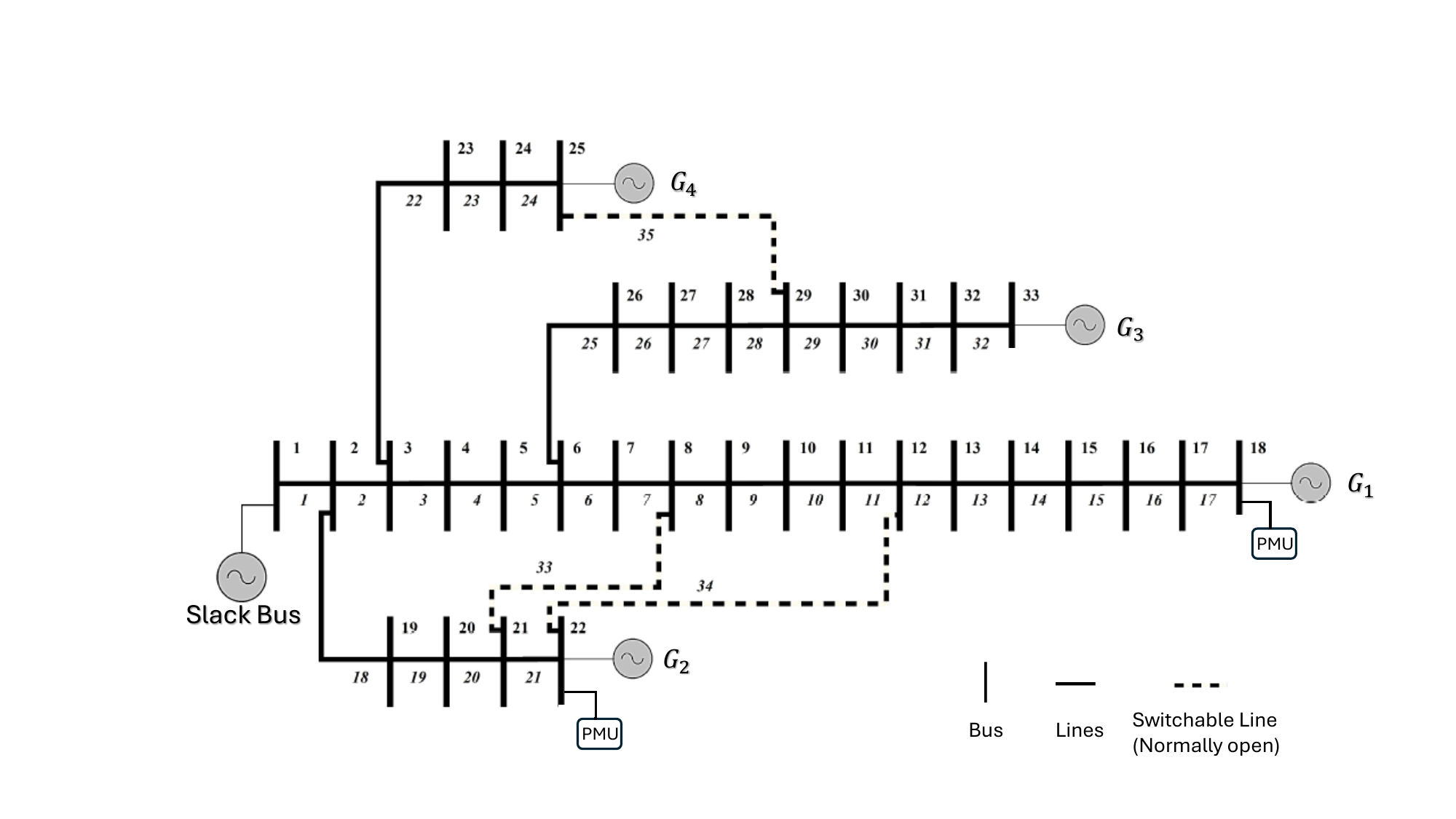}
        \centering
        \captionsetup{width=\linewidth}
        \caption{One line diagram of enhanced IEEE-33 bus system \cite{dolatabadi2020enhanced}}
        \label{IEEE33Bus}
\end{figure}

To evaluate the effectiveness of the LSHS method, we conduct simulations on a modified IEEE-33 bus system, as illustrated in Fig.~\ref{IEEE33Bus}. The detailed of the electric power system is provided in \cite{dolatabadi2020enhanced}.
The generator connected to bus 1 is the slack bus with a power capacity of 4 MW. Considering the states $x_i = [\delta_i, \omega_i]$ for each generator $G_i$, the state space used for the SHS model in Section~\ref{Sec3} is derived as $x = [x_1, \dots, x_4, \tilde{x}_1, \dots, \tilde{x}_4]^T$, where generators $G_1$ to $G_4$ each have an active power capacity of 0.2 MW. Additionally, two phasor measurement units (PMUs) are installed on buses 18 and 22 to measure the phasor angles of these two buses which are named $\delta_1$ and $\delta_2$, respectively. This measurement gains observability of the system. 
The line parameters for the system are derived based on \cite{dolatabadi2020enhanced}. 

The closed-loop SHS model has 16 states, with half of the state space representing the physical dynamics of the system, whose eigenvalues denoted as $\Lambda_1$, and the other half corresponding to the dynamics of the state estimation error, whose eigenvalues denoted as $\Lambda_2$.
 As described in Section III, physical contingencies influence the eigenvalues in both parts of the closed-loop SHS model. In contrast, control contingencies primarily affect $\Lambda_1$, and sensor/monitoring contingencies predominantly impact $\Lambda_2$. To illustrate the effects of contingencies on eigenvalues, we visualize the behavior of a physical contingency as a line outage on line 2 in Fig. \ref{eigenvalues}. Also, a control contingency is illustrated by increasing the control input $u_3$ of $G_3$ by 20\%, and a measurement contingency is shown as packet loss of $\delta_1$, as shown in Table. \ref{eigenvalues}.

\begin{figure}[t]
    \includegraphics[width=1\linewidth, trim={0cm 0cm 0cm 0cm},clip]{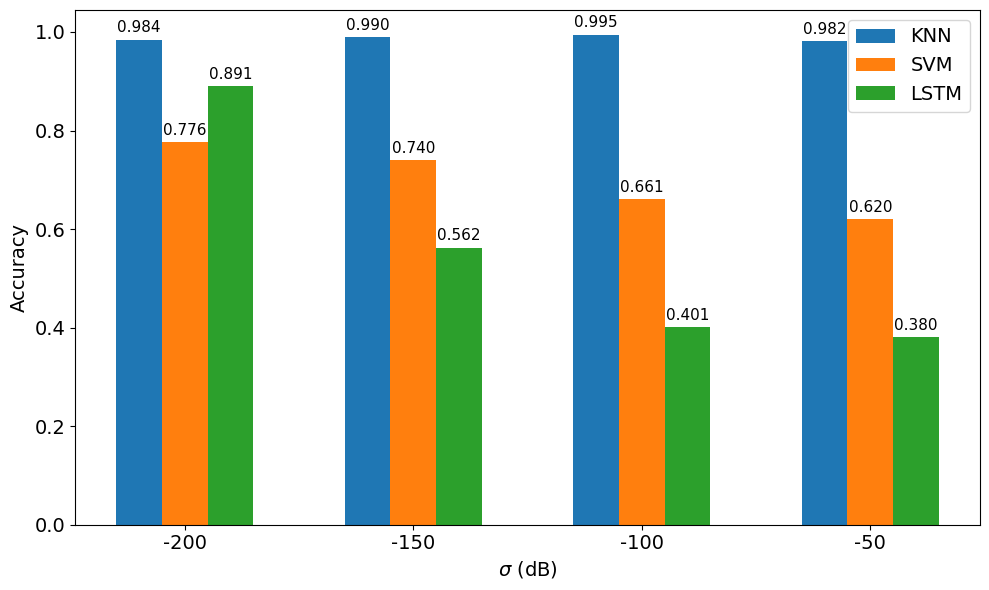}
        \centering
        \captionsetup{width=\linewidth}
        \caption{Performance Comparison of KNN, SVM, and LSTM Under Varying Noise Levels (\(\sigma\)).}
        \label{Accuracy}
\end{figure}

\begin{table}[t]
    \centering
    \caption{The closed-loop SHS model eigenvalues under normal operation, physical (line 2 outage), control (20\% increase of $u_3$), and measurement (packet loss in $\delta_1$) contingencies}
    \label{eigenvalues}
    \resizebox{\columnwidth}{!}{ 
    \begin{tabular}{|c|c|c|c|c|}
        \hline
        & Normal & Physical & Control & Measurement \\ 
        \hline
        \multirow{9}{*}{$\Lambda_1$} & $-1.096$ & $-1.003$ &  $-1.306$ & $-1.096$   \\ \cline{2-5}
        & $-0.833$ & $-0.848$ & $-0.833$ & $-0.833$  \\ \cline{2-5}
        & $-0.150$ & $-0.157$ & $-0.150$ &  $-0.150$ \\ \cline{2-5}
        & $-0.065$ & $-0.136$ & $-0.054$ & $-0.065$  \\ \cline{2-5}        
        & $-0.339 + j0.054$ & $-0.339 + j0.17$ & $-0.339 + j0.054$ & $-0.339 + j0.054$ \\ \cline{2-5}
        & $-0.339 - j0.054$ & $-0.339 - j0.17$ & $-0.339 - j0.054$ & $-0.339 - j0.054$ \\ \cline{2-5}
        & $-0.622 + j0.21$ &$ -0.622 + j1.41$ & $-0.622 + j0.21$ & $-0.622 + j0.21$  \\ \cline{2-5}
        & $-0.622 - j0.21$ & $-0.622 - j1.41$ & $-0.622 - j0.21$ & $-0.622 - j0.21$ \\ 
        \hline
        \multirow{9}{*}{$\Lambda_2$} & $-15$ & $-1.22$ & $-15$ & $-15$ \\ \cline{2-5}
        &$ -14$ & $-0.026$ & $ -14$ &  $-12$ \\ \cline{2-5}
        &$ -13$ & $0.4 + j0.939$ & $ -13$ &  $-9.005$ \\ \cline{2-5}
        & $-12$ & $0.4 - j0.939$ & $-12$ &  $-7.989$ \\ \cline{2-5}
        & $-11$ & $-0.219 + j0.15$ & $-11$  & $-0.067 + j0.662$ \\ \cline{2-5}
        & $-9$ & $-0.219 - j0.15$ & $-9$ & $-0.067 - j0.662$ \\ \cline{2-5}
        & $-8$ & $-0.001 + j0.01$ & $-8$ & $-0.067 + j0.662$ \\ \cline{2-5} 
        & $-7$ & $-0.001 - j0.01$ & $-7$ & $-0.067 - j0.662$ \\   \hline
    \end{tabular}
    } 
\end{table}
For training, the outputs of the closed-loop SHS is used as the dataset $y_c = [\delta_1, \delta_2, \hat{x}_1, \dots, \hat{x}_8]$. Samples are generated using the MATLAB linear state-space environment based on the SHS model parameters $t_s = 0.001$~s and $\tau_1 = 0.02$~s. The dataset consists of 960 scenarios, randomly generated with 240 samples for each class. For physical contingencies, we specifically consider a line outage scenario, where one of the lines is randomly disconnected. For control and monitoring contingencies, one input or sensor measurement value is randomly altered, ranging from zero to twice its actual value. The dataset is collected for different noise levels, 
$\sigma = \{-200dB, -150dB, -100dB, -50dB\}$
, by calculating the difference between the system output under contingency and the nominal output in normal operation without measurement noise. Labels for each data record are assigned accordingly. The time-series sequence from each scenario is used to train the LSTM algorithm, while the summation of the sequences is used to train the KNN and SVM algorithms. The accuracy of these algorithms at different noise levels is presented in Fig. \ref{Accuracy}. Accuracy is defined as the ratio of correct predictions to the total number of scenario samples. It is observed that the accuracy of SVM and LSTM decreases as noise levels increase, whereas the KNN algorithm remains robust to noise and outperforms the other two algorithms.

\begin{figure}[t]
    \includegraphics[width=0.8\linewidth, trim={0cm 0cm 0cm 0cm},clip]{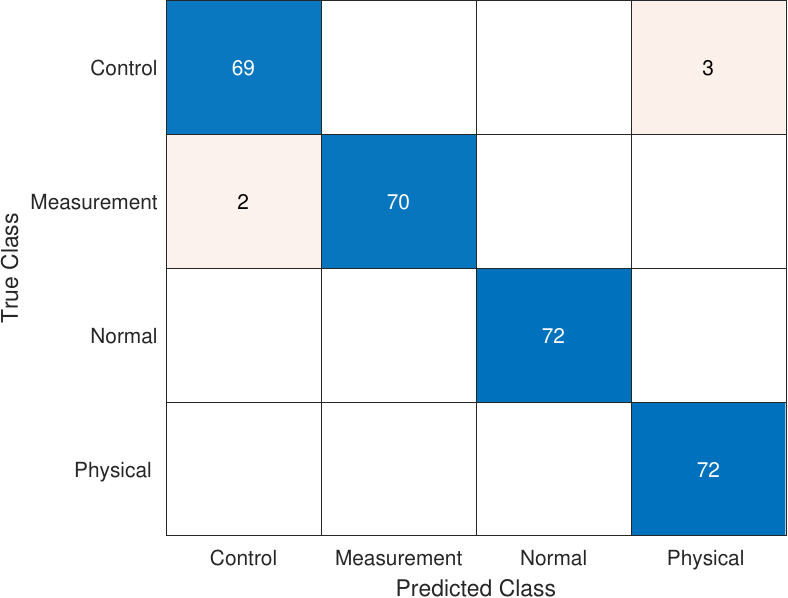}
        \centering
        \captionsetup{width=\linewidth}
        \caption{Performance of the KNN algorithm for Training Dataset with K=1 and Accuracy = 98.29\%.}
        \label{KNNconfusion}
\end{figure}

Based on its superior performance, the K-nearest neighbors (KNN) algorithm has been selected for implementation in the LSHS framework. The confusion matrix of the KNN algorithm is shown in Fig. \ref{KNNconfusion}, demonstrating an accuracy of $98.29\%$ with $k=1$. 

This system will be utilized for detecting the occurrence of contingencies and classifying them into their respective categories. Furthermore, it will identify the exact contingency scenario, enabling a direct performance comparison between the proposed LSHS framework and the method introduced in \cite{cd2}. This comprehensive evaluation highlights the effectiveness and robustness of the LSHS approach.

\begin{figure}[t]
    \includegraphics[width=1\linewidth, trim={0cm 0cm 0cm 0cm},clip]{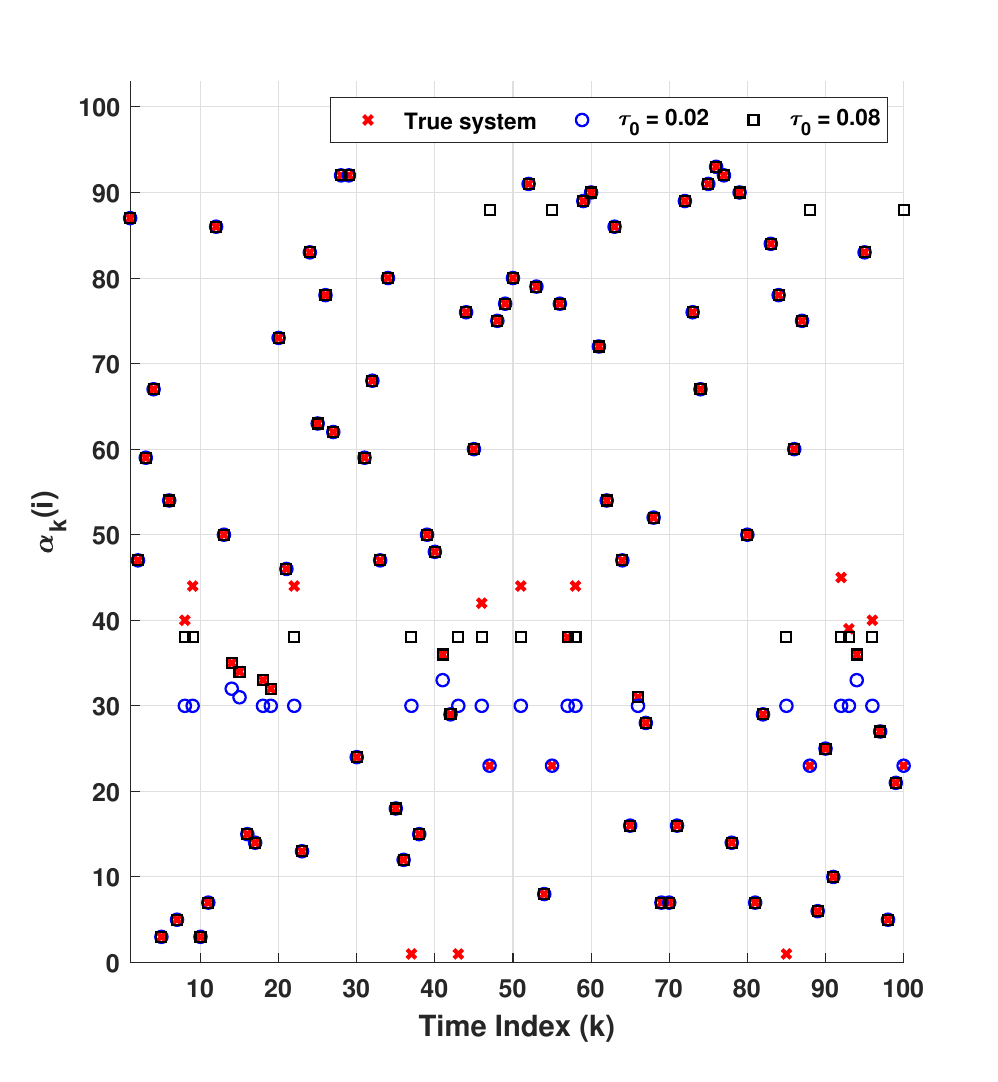}
        \centering
        \captionsetup{width=\linewidth}
        \caption{Random switching of the MPS and identification of contingencies by the SHS algorithm in \cite{cd2}.}
        \label{CD2switching}
\end{figure}

\begin{figure}[t]
    \includegraphics[width=1\linewidth, trim={0cm 0cm 0cm 0cm},clip]{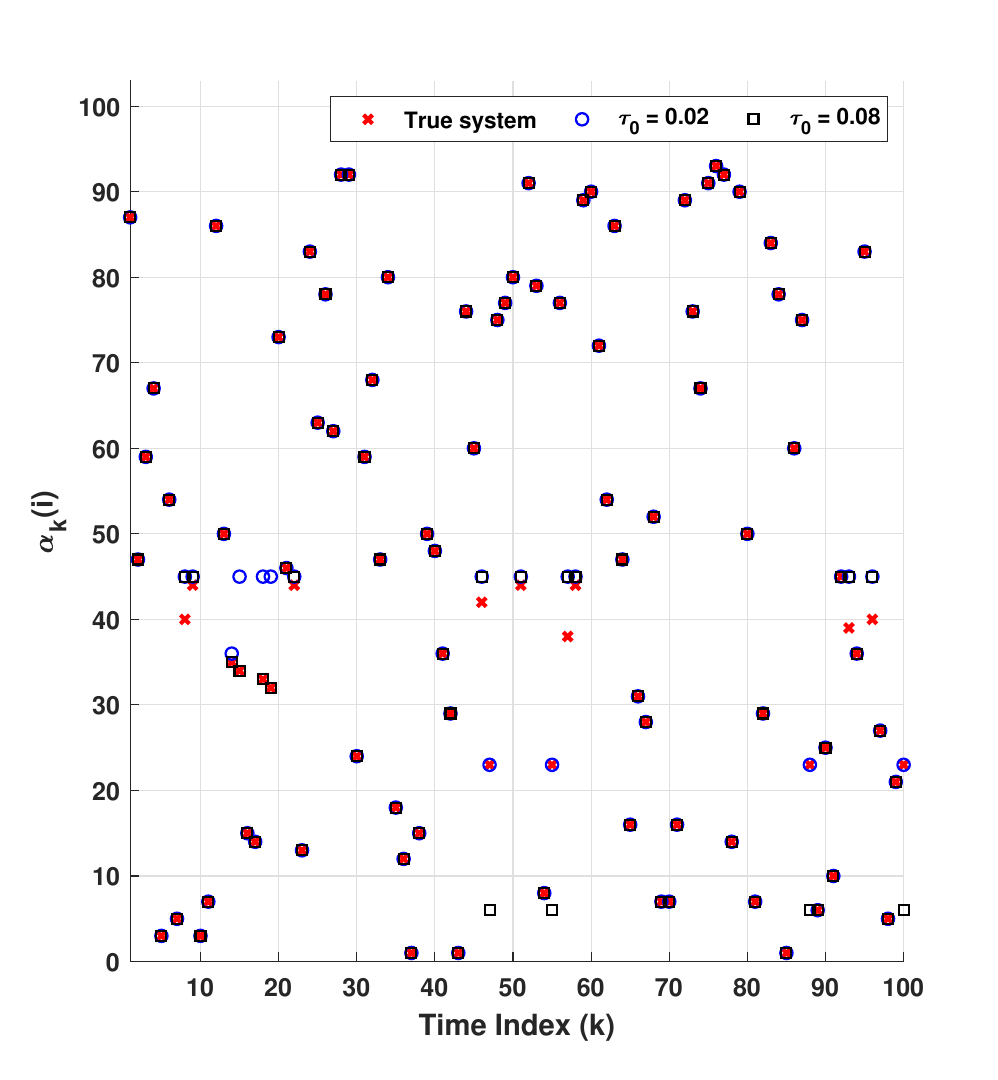}
        \centering
        \captionsetup{width=\linewidth}
        \caption{Random switching of the MPS and identification of contingencies after applying LSHS.}
        \label{AMPSswitching}
\end{figure}

For the IEEE-33 bus system, the LSHS contingency dataset is defined as follows: \(\alpha_k = 1\) represents normal operation. \(\alpha_k = 2\) to \(29\) correspond to $N-1$ line outage scenarios. Control contingencies are represented by \(\alpha_k = 30\) to \(61\), characterized by 25\%, 50\%, 75\%, and 100\% increases and decreases in the control input of generators. Sensor/monitoring contingencies are defined by \(\alpha_k = 62\) to \(93\), which include increases and decreases in measurements by \(\pm80\%\) in steps of 10\%. Then, system  operates under 500 sequences of random switching as the contingency scenarios. We applied the contingency identification algorithm of \cite{cd2} for three different values of \(\tau_0 = 0.02\), \(0.05\), and \(0.08\), where \(\tau_1 = 0.02\). For more clarity, the results are depicted in Fig. \ref{CD2switching} and Fig. \ref{AMPSswitching}, illustrating 100 sequences of random switching for \(\tau_0 = 0.02\) and \(\tau_0 = 0.08\).
Higher ability of LSHS on identifying the contingencies is observable compared to identification based on the SHS approach only.
In addition, for the control contingencies (\(\alpha_k = 30\) to \(61\)), neither algorithm demonstrates satisfactory performance, as they fail to detect these contingencies completely.

\begin{figure}[t]
    \includegraphics[width=1\linewidth, trim={0cm 0cm 0cm 0cm},clip]{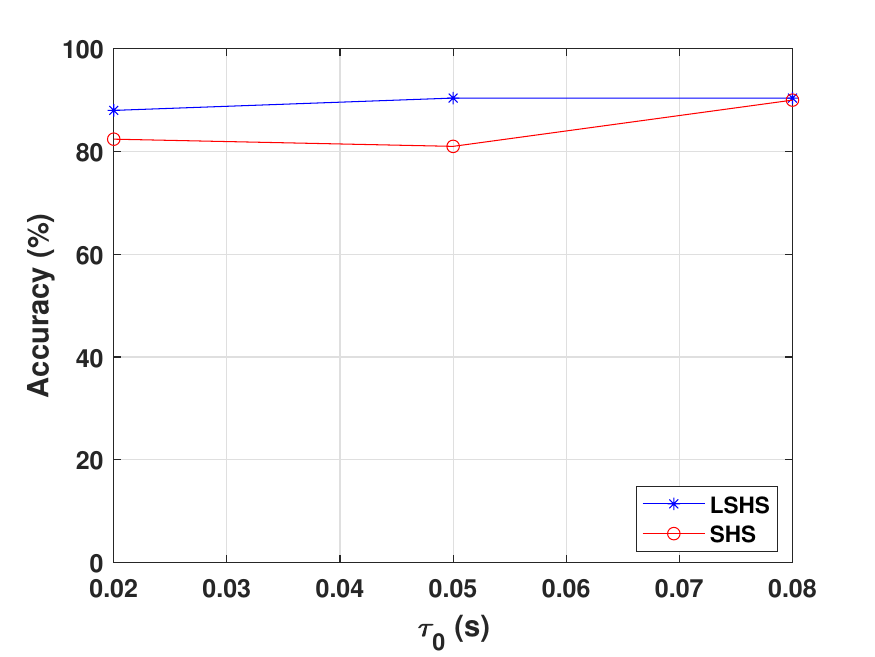}
        \centering
        \captionsetup{width=\linewidth}
        \caption{Comparing the accuracy of contingency identification for LSHS and SHS methods for different values of $\tau_0$.}
        \label{Accuracies}
\end{figure}

Next, we define an accuracy metric, which represents the ratio of correctly detected contingencies to the total number of contingencies. As shown in Fig.~\ref{Accuracies}, the accuracy is plotted for three different values of \(\tau_0 = 0.02\), \(0.05\), and \(0.08\). The results demonstrate that LSHS is capable of achieving better performance compared to SHS at smaller values of \(\tau_0\). This is particularly evident when \(\tau_0 = 0.05\). The other advantage of using LSHS is that it can detect contingencies within a shorter time interval ($\tau_1$). This increase the speeed of contingency detection. 

\begin{figure}[t]
    \includegraphics[width=1\linewidth, trim={0cm 0cm 0cm 0cm},clip]{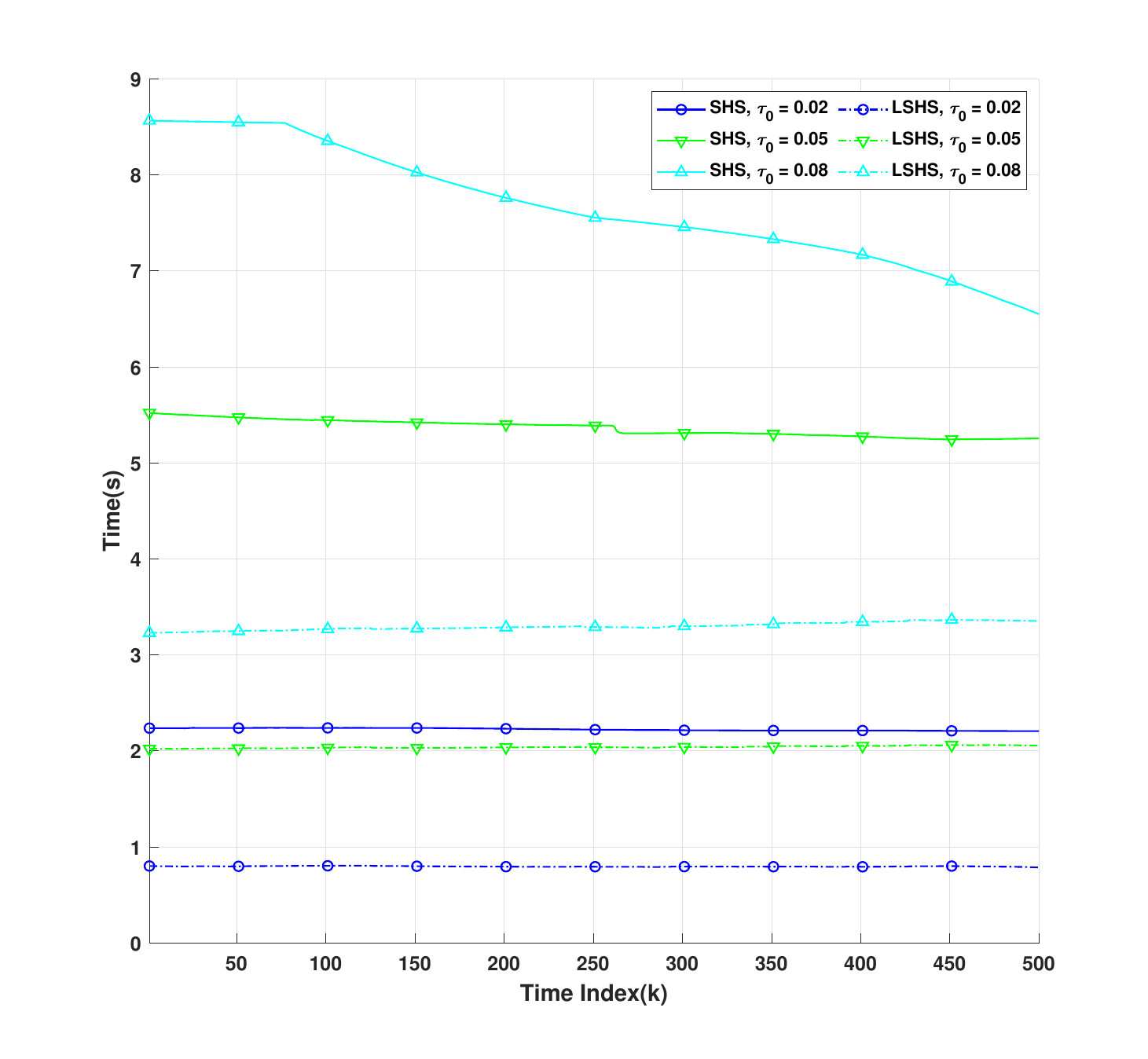}
        \centering
        \captionsetup{width=\linewidth}
        \caption{Comparing the moving average of the time spent for contingency identification for the LSHS and SHS methods with different values of $\tau_0$.}
        \label{TimeAverage}
\end{figure}

Furthermore, Fig.~\ref{TimeAverage} highlights the moving average of the time spent for contingency identification. In other words, LSHS not only enhances accuracy but also reduces the identification time. For each $\tau_0$, LSHS requires less time for identification compared to the SHS algorithm in \cite{cd2}.
  
\section{Conclusion}
\label{Sec6}

This paper presents a novel learning-based stochastic hybrid system for contingency detection and classification in MPS. By categorizing contingencies into physical, control network, and measurement classes, the LSHS method provides deeper insights into their impact on system dynamics and state estimation errors.

The study demonstrates several advantages of applying the LSHS algorithm prior to contingency identification: 1) Faster Detection of Hidden Contingencies:
The LSHS algorithm enables the early detection of hidden contingencies, by analyzing the system output behavior and without needs for direct measurement of the contingency. This eliminates the need for continuous probing input implementation during normal operation, significantly improving both the speed and accuracy of contingency detection. 2) Efficient Classification and Reduced Search Space:
By classifying contingencies into physical, control, and sensor/monitoring categories, the LSHS method narrows the search space for contingency identification. This results in two key benefits:
i) Higher accuracy in shorter time frames, which results in faster identification with improved precision.
ii) Reduced Computational burden, where system response estimation is required only for the relevant class of contingencies, minimizing unnecessary calculations.

The effectiveness of the proposed LSHS method is validated through simulations on a modified IEEE-33 bus system. Results show that LSHS can detect all types of contingencies within a short time interval (0.02 seconds) with an accuracy exceeding 98\%. These findings highlight the robustness and efficiency of the LSHS framework for enhancing the reliability and resilience of MPS. Future research will focus on leveraging hidden Markov models and deep learning techniques to monitor and assess the risk of cascading chain reactions in MPS. This will enhance predictive capabilities and system resilience.

\bibliographystyle{ieeetr}
\bibliography{references}

\end{document}